\documentclass{article}

\usepackage{arxiv}

\usepackage[utf8]{inputenc} % allow utf-8 input
\usepackage[T1]{fontenc}    % use 8-bit T1 fonts
\usepackage{hyperref}       % hyperlinks
\usepackage{url}            % simple URL typesetting
\usepackage{booktabs}       % professional-quality tables
\usepackage{amsfonts}       % blackboard math symbols
\usepackage{nicefrac}       % compact symbols for 1/2, etc.
\usepackage{microtype}      % microtypography
\usepackage{graphicx}
\usepackage{epstopdf, epsfig}
\usepackage{amsmath}
\usepackage{graphicx}
\usepackage{graphics}
\usepackage{subfig}
\usepackage{listings}

\title{Deep Reinforcement Learning in Fluid Mechanics: a promising method for both Active Flow Control and Shape Optimization}
\author{
  Jean Rabault\\
  Department of Mathematics\\
  University of Oslo\\
  \textit{corresp. author: jean.rblt@gmail.com} \\
   \And
  Feng Ren \\
  Research Center for Fluid Structure Interactions\\
  Department of Mechanical Engineering \\
  Hong Kong Polytechnic University \\
   \And
  Wei Zhang \\
  Science and Technology on Water Jet Propulsion Laboratory, \\
  Marine and Research Institute of China, \\
  Shanghai
   \And
  Hui Tang \\
  Research Center for Fluid Structure Interactions\\
  Department of Mechanical Engineering \\
  Hong Kong Polytechnic University \\
   \And
  Hui Xu \\
  School of Aeronautics \& Astronautics, \\
  Shanghai Jiao Tong University \\
}

\begin{document}
\maketitle

\begin{abstract}
In recent years, Artificial Neural Networks (ANNs) and Deep Learning have
become increasingly popular across a wide range of scientific and technical
fields, including Fluid Mechanics. While it will take time to fully grasp
the potentialities as well as the limitations of these methods, evidence is
starting to accumulate that point to their potential in helping solve
problems for which no theoretically optimal solution method is known. This
is particularly true in Fluid Mechanics, where problems involving optimal
control and optimal design are involved. Indeed, such problems are famously
difficult to solve effectively with traditional methods due to the
combination of non linearity, non convexity, and high dimensionality they
involve. By contrast, Deep Reinforcement Learning (DRL), a method of
optimization based on teaching empirical strategies to an ANN through trial
and error, is well adapted to solving such problems. In this short review,
we offer an insight into the current state of the art of the use of DRL
within fluid mechanics, focusing on control and optimal design problems.
\end{abstract}

\section{Introduction}

The application of Machine Learning (ML) to classical scientific and technical fields of research is
advancing rapidly. One key factor of the successes of ML is the development of Artificial
Neural Networks (ANNs), which approximate a complex function using a collection of connected
units called neurons, where signals between neurons are processed using simple mathematical operations.
This results in a method that is able to deal with a variety of
problems, ranging from image analysis \citep{krizhevsky2012imagenet, he2016deep, rabault2017performing} to the optimal control of robots
\citep{doi:10.1177/0278364913495721, 7989385}. These recent successes have highlighted the ability of ANNs to deal with
the ingredients of non linearity, non convexity, and high dimensionality that
are still today a challenge across a wide range of scientific disciplines \citep{lusch2018deep}.

Following the simple fully connected ANN concept, more complex network architectures have been proposed, for example, the
Convolutional Neural Networks (CNNs) that are advantageous in image recognition and the Recurrent Neural
Networks (RNNs) that are advantageous in natural language processing. These neural network
architectures, which feature a combination of multiple types of layers, build up the base of the Deep Learning (DL) revolution.
Another significant branch of the ML is the Deep Reinforcement Learning (DRL) field, in which an agent
seeks optimal actions to interact with an environment, so as to maximize a cumulative reward.
Taking the proximal policy optimization (PPO) algorithm \citep{heess2017emergence, Schulman2017}
for example, the DRL usually adopts two sets of deep neural networks, i.e., one to approximate the
reward, and the other to build the relationship between states of the environment and the optimal
actions. Successful applications of the DRL includes the well-known AlphaGo that defeated the top-level
human player at the game of Go \citep{silver2017mastering}, legged robots deployed in real environments
\citep{hwangbo2019learning}, and many other applications. Therefore, these results motivate researchers from the Fluid Mechanics
community to explore the combination of the DRL with Fluid Mechanics in areas such as biomimetic
birds flight \citep{Learning_to_soar_in_turbulent_environments}, or fish swimming \citep{verma2018efficient}.

In the present paper, we aim at providing a concise review of
Deep Reinforcement Learning (DRL, \citep{SCHMIDHUBER201585, Lillicrap2015, Sutton2018}), focusing on its use to solve problems
arising in Fluid Mechanics. Our aim here is to provide a general overview of
the field, point the reader to the relevant literature for more details, and
present the main challenges that remain to be solved for further application of
this method. This short review was written as a complement to the presentation
of the same title held at the International Symposium on High-Fidelity
Computational Methods and Applications in Shanghai, December 2019 \citep{shanghaiRBTpres},
and it also shares a lot of material and ideas with the lecture held at the 
2019 Flow/Interface School on Machine Learning and Data Driven Methods,
KTH, Stockhom 2019 \citep{KTHWinterSchool} (the slides of this extended lecture are available
at \url{https://folk.uio.no/jeanra/Research/material_lecture_DRL_FM_Rabault_2019.pdf}). The
aim of this review is therefore to offer a self-contained summary of the
material presented there.

The organisation of the paper is as follows. First, a brief introduction to
both ANNs and DRL is provided. Then, the application of DRL to both optimal
control and optimal design is presented. Finally, we discuss future prospects
and offer some words of conclusion.

\section{A brief introduction to Artificial Neural Networks and Deep Reinforcement Learning}

\subsection{Artificial Neural Networks and supervised learning}

\begin{figure*}[ht]
\begin{center}
\includegraphics[width=.45\textwidth]{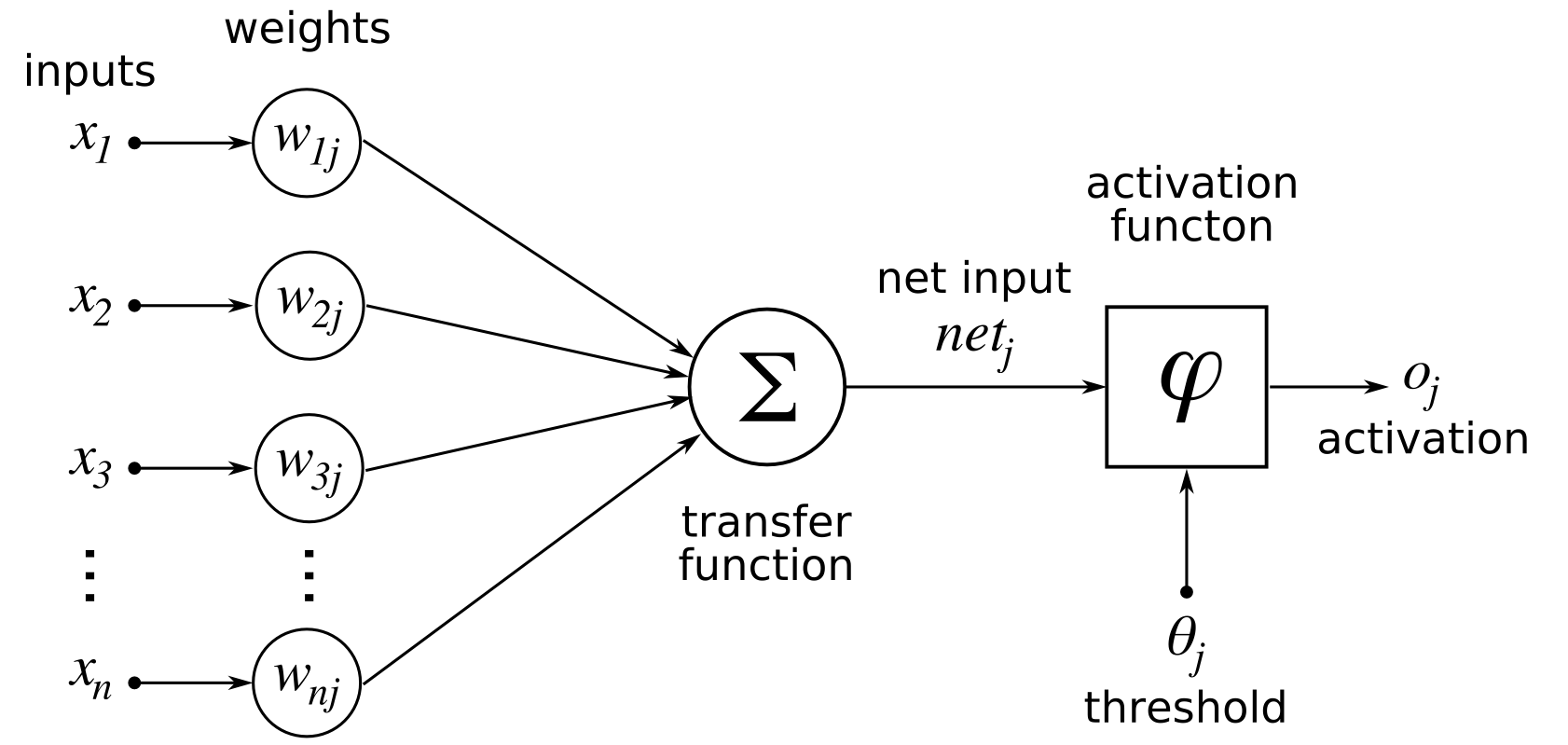}
\includegraphics[width=.45\textwidth]{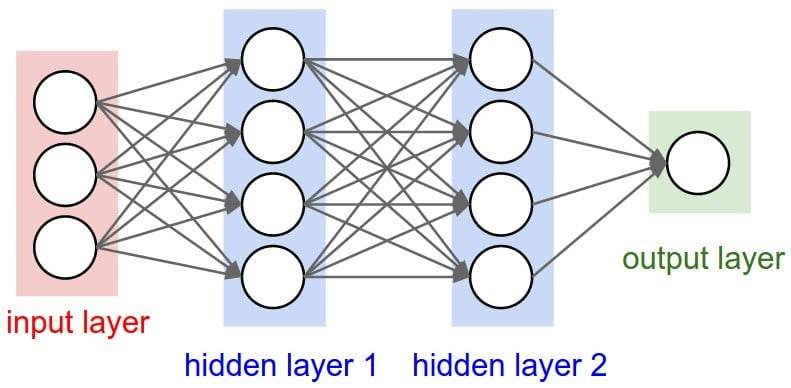}
    \caption{\label{fig:NeuronAndANN} Illustration of the functioning of an
    artificial neuron (left). Neurons can be organized in layers to define a fully
    Artificial Neural Network (ANN) (right).
}
\end{center}
\end{figure*}

ANNs are based on ideas proposed as early as the 1950s \citep{rosenblatt1957perceptron}, which take their
inspiration in a loose analogy to the functioning of the cortex of rats.
More specifically, an ANN is defined as a set of neurons, where each neuron is
a simple computational unit \citep{DeepLearningLeCunNature, Goodfellow2017}. In general, only feedforward (i.e. acyclic)
ANNs are considered, and a neuron is composed of two parts. First, it collects
information from a series of inputs and performs a weighted sum of these, and
second, it applies an activation function to the result obtained at the first
stage. In this context, the set of weights of all the neurons of the network
defines the parametrization of the model. Usually, an ANN is composed of a
series of layers of neurons, where the outputs of all neurons of a layer are
used as inputs to the next layer. This is summarized in Fig.
\ref{fig:NeuronAndANN}. In addition, specific architectures can be used to
reflect the structure of a given problem. The most famous example there is the
Convolution layer \citep{lecun1995convolutional}, which considers the input to the layer as a
N-dimensional structured data and generates as output a similarly structured
data by application of a convolutional kernel on the input, following a regular
pattern. Therefore, Convolutional Neural Networks (CNNs) effectively apply the same local transformation on
smaller parts of their input, which enforces both locality and translational
invariance of the processing they perform.

From a formal mathematical point of view, the interest of ANNs lies in their
property of being universal approximators \citep{Hornik1989359}. What is meant there, is that
an ANN using a suitable nonlinear activation function can, given that its size
is large enough, perform an arbitrarily good approximation of any function from
its input to its output layers. However, while this result holds in theory, in
practice there are many challenges to performing effective learning. First,
this theoretical result says nothing about which exact activation function to
use, or how large the network should be. Therefore, a large body of literature
has discussed both these questions. In practice, it is commonly agreed nowadays
that a simple activation function, such as the Rectified Linear Unit (ReLU), is
often a good choice \citep{AISTATS2011_GlorotBB11}. Similarly, the literature reports that the depth of
ANNs is important for their efficiency and descriptive power \citep{DeepLearningLeCunNature}. Therefore,
recent state of the art ANNs can feature up to over a hundred layers \citep{he2016deep}. However,
this increased depth comes with many challenges, in particular the risk for the
signal propagating in the ANN to either explode or decay exponentially with
increasing depth \citep{Glorot10understandingthe}. As a consequence, both regularization techniques
(dropouts \citep{srivastava2014dropout}, batch normalization \citep{ioffe2015batch}), specific network initializations
(Xavier initialization \citep{Glorot10understandingthe}, He initialization \citep{he2015delving}), and specific architectures (Resnets \citep{he2016deep}, UNet \citep{li2018h}) have been developed to help solve these challenges. These
technical refinements, together with the use of backpropagation of errors as a
training method \citep{DeepLearningLeCunNature}, and some effective batch optimizers \citep{duchi2011adaptive, kingma2014adam}, are now the
most common methodology used to train ANNs to model a given function or
dataset. Following these advances, ANNs have now reached the point where, at
least from the point of view of most practitioners, supervised learning, i.e.
learning a function from a large database of labeled examples, works
satisfactorily on many tasks such as image classification or speech analysis as
long as enough data and computational power are available.

The use of ANNs is made especially easy thanks to the development of several
high quality open source implementations. These include, to name but a few,
Tensorflow \citep{abadi2016tensorflow} (a framework developed largely by Google), and Torch /
pyTorch \citep{NEURIPS2019_9015} (developed by Facebook). Thanks to these frameworks, using ANNs
has become very easy and the practitioners can focus purely on their problem,
rather than in the low-level implementation of all the technical features of
modern ANNs.

\subsection{Deep Reinforcement Learning, Policy Gradient methods, and the Proximal Policy Optimization algorithm}

\begin{figure*}[ht]
\begin{center}
\includegraphics[width=.45\textwidth]{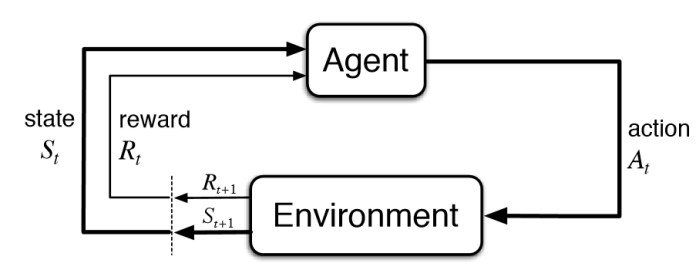}
    \caption{\label{fig:DRLFramework} Illustration of the DRL framework. A
    closed control loop is defined through the interaction between the
    environment and the agent through the state and action channels of
    communication, and the reward signal is used to drive the optimization
    process. For this, an ANN is used as a function approximator and
    trained to optimize the reward.
}
\end{center}
\end{figure*}

As a consequence of this success, ANNs and supervised learning can also be used
as building blocks in more sophisticated algorithms. One such field of
application that is encountering large success recently is the Deep
Reinforcement Learning (DRL) methodology \citep{Sutton2018}. In Deep Reinforcement Learning,
the ANN is used as a function approximator being part of a larger algorithm
that performs learning through trial-and-error. For this, a formal
framework is defined encompassing on the one hand the agent, i.e. the ANN
together with algorithms that allow to perform the trials-and-errors and
learning, and on the other hand the environment, i.e. the system to control.
The interaction between the agent and the environment takes place through 3
standardized channels of interaction: a state observation $s$ that can be
noisy, partial, and stochastic, an action $a$ that is the control applied on
the environment, and a reward signal $r$ provided by the environment to the
agent that indicates how good the current state of the environment is. This
interaction naturally takes place in a closed-loop fashion, and the aim of DRL
is to maximize the obtained reward by judicious choice at each step of the
control, given the state information. The general DRL framework is illustrated
in Fig. \ref{fig:DRLFramework}. Several striking, high-profile successes have
been obtained using DRL and its refinements in recent years, such as winning at
the game of Go against the best human player \citep{silver2017mastering, Silver1140}, or effectively controlling
server farms to reduce cooling electricity consumption \citep{GoogleDataCenter2018}.

Different algorithms can be used to perform the learning through
trial-and-error. Most of them fall in two large categories, namely Q-learning
which is based on the Bellman equation \citep{Bellman1957, Bellman1962, DDQN}, and Policy Gradient (PG) \citep{Schulman2015}. Both categories contain many
algorithmic variants and refinements \citep{garnier2019review}. In the following, we will focus mostly on
PG, which has provided several algorithms regarded as state-of-the-art for
controlling processes where a continuous action space is present.

In the general framework of the policy gradient method, the aim is to optimize
the discounted reward $\sum_{t \geq 0} \gamma^t r_t$, where $\gamma$ is a
discount factor between 0 and 1, typical $\gamma = 0.95$, and $t$ indicates
timestep number. For a value of $\gamma$ close to 0, only the current reward is
important, while for $\gamma=1$ the agent does not prefer to obtain the reward
at the current time step or later in time, therefore favoring optimization over
a distant time horizon. The optimization takes place with regards to the set of
weights $\Theta$ of the ANN, where the ANN is learning a mapping from state
inputs to actions. The actions are described through probability density
functions (pdfs), which are parametrized by the outputs of the network.
Depending on the nature of the output (continuous, discrete, with compact
support or not), different parametric pdfs can be used. A common choice for
continuous output on compact support is for example to resort to a scaled beta
distribution.

In order to formulate the optimization problem as a gradient descent, it is
convenient at this point to introduce the concept of trajectory in the phase
space, $\tau$. A such trajectory is a sequence of pairs of state and actions,
$\tau = (s, a)_t$, where $t=0..N$ is the step in the current trajectory.
Following this definition, the value function to optimize can be written as:

\begin{equation}
V(\Theta) = \mathbb{E} \left[ \sum_{t=0}^{H} R(s_t, a_t) | \pi_{\theta} \right] = \sum_{\tau} \mathbb{P}(\tau, \Theta) R(\tau),
\end{equation}

\noindent where $\pi_{\theta}(a | s)$ is the policy, i.e. the probability
of taking action $a$ given the state observation $s$. The policy is parametrized
by an ANN with set of weights $\Theta$.

Therefore, one can take the gradient of this expression, and apply simple
manipulations to obtain:

\begin{align}
    \nabla_{\Theta} V(\Theta) &= \sum_{\tau} \nabla_{\Theta} \mathbb{P}(\tau, \Theta) R(\tau) \\
                              &= \sum_{\tau} \frac{\mathbb{P}(\tau, \Theta)}{\mathbb{P}(\tau, \Theta)} \nabla_{\Theta} \mathbb{P}(\tau, \Theta) R(\tau) \\
                              &= \sum_{\tau} \mathbb{P}(\tau, \Theta) \nabla_{\Theta} \log \left( \mathbb{P}(\tau, \Theta) \right) R(\tau).
  \end{align}

The last formula can be interpreted as a Monte Carlo formula, and estimated
through sampling of enough trajectories in the phase space:

\begin{equation}
 \nabla_{\Theta} V(\Theta) \approx \hat{g} = \frac{1}{M} \sum_{i=1}^{M} \nabla_{\Theta} \log \left( \mathbb{P}(\tau^{(i)}, \Theta) \right) R(\tau^{(i)}).
 \end{equation}

The physical interpretation is that the gradient estimated modifies the weights
of the ANN so as to increase the probability of trajectories that have a high
reward, and reduce the probability of trajectories that have a low reward. The
gradient of the log probability can be derived following:

\begin{align}
\nabla_{\Theta} \log \left( \mathbb{P}(\tau^{(i)}, \Theta) \right) &= \nabla_{\Theta} \log \left[ \prod_{t=0}^{H} \mathbb{P}(s_{t+1}^{(i)} | s_{t}^{(i)}, a_t^{(i)}) \pi_{\tau}(a_t^{(i)} | s_{t}^{(i)})  \right] \\
                                                                       &= \nabla_{\Theta} \left[ \sum_{t=0}^{H} \log \mathbb{P}(s_{t+1}^{(i)} | s_{t}^{(i)}, a_t^{(i)}) + \sum_{t=0}^{H} \log \pi_{\tau}(a_t^{(i)} | s_{t}^{(i)})  \right] \\
                                                                       &= \nabla_{\Theta} \sum_{t=0}^{H} \log \pi_{\tau}(a_t^{(i)} | s_{t}^{(i)}),
  \end{align}

\noindent where we have used the fact that the transition probabilities are
only functions of the environment and its dynamics, and are therefore not
related to the weights of the ANN. The gradient of the policy can be directly
related to the weights of the ANN through backpropagation of errors,
remembering that the output of the ANN is used to parametrize the distribution
from which actions are sampled.

Therefore, combining the formula (5) and (8) provides an algorithm for
performing DRL following the Policy Gradient method. Usually, deploying
such algorithms takes place in two phases. First, during training, one can
sample trajectories in the phase space under the policy provided by the ANN, by
randomly sampling the pdf provided by the ANN at each time step. This,
therefore, naturally introduces stochasticity in the trajectories taken and
drives exploration. Second, once training has been performed, one can follow
the optimal policy by choosing as action at each time step the peak of the pdf
produced by the ANN. This exploitation of the ANN to produce an
optimal policy (or rather, `believed to be' optimal policy) is referred to in the literature as `deterministic mode', or
`single runner mode'.

While in theory the method presented here could be applied as is, in practice a
number of refinements are used to make the convergence more robust and stable.
These refinements can include, depending on the exact DRL algorithm, the use of
a memory replay buffer \citep{PrioritizedER}, the use of a separate network (critic network) to
learn an estimator of the actualized reward \citep{Pinto2017}, the imposition of
limitations to the policy updates to avoid overfitting lucky events \citep{Schulman2015, Schulman2017}, and
corrections to the formula (5) and (8) following importance sampling to allow
the use of slightly off-policy data while training, therefore improving sample
efficiency. Such improvements can be combined to formulate more efficient
methods, such as the Proximal Policy Optimization (PPO) \citep{Schulman2017}, which is
currently often regarded as the state-of-the-art algorithm for performing
control of continuous action domain processes.

In addition to these low-level algorithmic improvements, high-level
improvements can also be implemented. For example, one can enhance the
exploration abilities of the PPO by implementing surprise or curiosity mechanisms \citep{achiam2017surprise}.
For this, the ANN is asked to not only choose the next control value, but also
to predict what the expected state will be after the control has been applied.
Thereafter, the reward is enriched as $r'_t = r_t + s_t$, where $s_t$ is the
surprise value, i.e. a measure of the discrepancy between the state effectively
reached and the state predicted by the ANN. As a consequence, the ANN builds an
intrinsic reward that drives it to prioritize exploring trajectories on which
the prediction of the next state following control is of bad quality. Such a
region where the prediction by the ANN is in bad agreement with the next state
indicates that either the corresponding region was not or poorly explored, or
that complex dynamics not yet understood by the ANN take place in this domain
of the phase space. Therefore, the curiosity-driven ANN is pushing itself to
investigate in more details regions of the phase space where its knowledge is
inaccurate, which incentivizes effective exploration. Unfortunately, this
can lead to the ANN being stuck exploring again and again regions of true
randomness, similar to the cognitive phenomenon of procrastination in human
psychology \citep{savinov2018episodic}. Therefore, more sophisticated curiosity mechanisms have been
developed, such as curiosity through reachability \citep{savinov2018episodic}, that preserve the desirable
properties of curiosity mechanisms while alleviating negative aspects such as
procrastination. Curiosity is not the only way to enhance learning with
high-level techniques. For example, authors have reported that resorting to
World Models \citep{ha2018world}, i.e. forcing an ANN to learn a dynamic model of the
environment and training subsequently on it in addition to the main environment
(an idea which is inspired by the mechanisms of dreaming in human psychology),
can also improve learning. Finally, one can also take advantage of human expert
knowledge, and improve the initial phases of training through showing examples
to the ANN, or by carefully crafting which initial state the ANN starts to
control the system from \citep{salimans2018learning}. This in turn can reduce the number of steps
necessary to reach a desirable state of the system, effectively cutting on the
cost of exploration.

At this stage it should be pointed out that, while an intrinsically local
gradient descent approach is used to train the ANN, the method as a whole
performs, by contrast, a global optimization that is well adapted to fully non
linear, non convex, high dimensionality systems. Indeed, the true mechanism
behind the collection of data and therefore the discovery of new strategies
lies in the collection of trajectories in the phase space performed during
training, which presents a large stochasticity. Therefore, given that enough
trajectories are sampled, the PPO algorithm has the property to explore the
behavior of the environment even past low reward barriers (which are the
analogy for DRL to high potential energy barriers in classical physics).

As a consequence of these strengths the PPO algorithm (or similar DRL / PG algorithms),
together with large amounts of computational power to allow the sampling of
a large number of trajectories in the phase space, is at the origin of several
successes of the DRL approach on complex systems. While there is little
interest to use DRL on systems where an optimal, analytical or semi-analytical
control algorithm is known, many realistic problems of interest cannot be
solved effectively using such traditional approaches in which case DRL and PPO
are natural methods to investigate. This efficiency is not specific to a given
field of research, which is clearly reflected in the literature since DRL
methods have been found to be successful in a wide variety of domains. In
the following, we will focus on applications to Fluid Mechanics, and more
specifically to optimal flow control and optimal design.

Before presenting applications of DRL to fluid mechanics we want to mention
that, similarly to what was described for supervised learning, high quality
open source implementations of the most recent DRL algorithms such as PPO are
available freely. Such implementations include, to name but a few, Tensorforce
\citep{tensorforce}, and stable-baselines \citep{stable-baselines}. Both packages implement a wide
variety of Q-learning and Policy Gradient methods and provide a number of
helper functions to make deployment easier. In addition, these libraries are
built on top of tensorflow, which makes them computationally effective. From
the point of view of the practitioner, all what is needed to use these packages
is to implement the application-specific environment. For this, it is enough to
fill-in templated interface classes. For example, tensorforce provides an
environment class from which custom environments should be derived. The main
structure of the custom environment class should be as follows:

\begin{lstlisting}[language=Python]
from tensorforce.environments import Environment

class MyEnvironment(Environment):
    def states(self):
        """Returns the state space specification.
        """

    def actions(self):
        """Returns the action space specification.
        """

    def reset(self):
        """Resets the environment to start a new episode.
        Returns: dict[state]: Dictionary containing initial
            state(s) and auxiliary information.
        """

    def execute(self, actions):
        """Executes the given action(s) and advances the
        environment by one step.
        Args: actions (dict[action]): Dictionary containing
            action(s) to be executed
        Returns:
            ((dict[state], bool | 0 | 1 | 2, float)):
            Dictionary containing next state(s),
            terminal information, and observed reward.
        """
\end{lstlisting}

The structure of this class closely follows the DRL framework presented in Fig.
\ref{fig:DRLFramework}. At the start of each episode (i.e. trajectory in the
phase space), the \textit{reset} method is called to initialize the environment
and derive the initial state. Following this initialization, the
\textit{execute} method can be called as many time as requested to enforce the
closed-loop interaction. Finally, a number of additional functions and example
scripts are provided to the user on for example the Tensorforce repository
\url{https://github.com/tensorforce/tensorforce}, to help perform both training
and single run exploitation. For a complete overview, see either the tensorforce
documentation \url{https://tensorforce.readthedocs.io/en/latest/}, or one of the active flow control codes released as open
source \url{https://github.com/jerabaul29/Cylinder2DFlowControlDRL}.

\section{Applications to control}

Control of intermediate to high Reynolds number flows is a natural field of
application for DRL. Indeed, the complexity of such flows is usually too high
for traditional mathematical methods to provide positive results and,
therefore, the field of flow control is famously difficult to apprehend with
analytical tools \citep{duriez2017machine}. As a consequence, two main applications of DRL can be
found there. The first one is the high-level control of large systems, such as
the swimming of a fish-like geometry, or the flying of a bird-like object. The
second one is the finer grained, more detailed active flow control at the small
scale of flow instabilities and separation.

Arguably, this first domain of application (i.e., the high-level control of
large systems) is at the interface between robotics, biomimetism, and fluid
mechanics. Several high-profile works have been published on this topic
recently. To name but a few, the swimming of fish schoolings has been
investigated through simulations and revealed how the vortex alley generated by
a leader fish can be exploited by a follower fish to reduce energy consumption
and increase swimming speed \citep{Novati2017}. This, in turn, may find applications for robotic
swarms. Another high-profile application of DRL to high-level control problems
can be found in \citet{Learning_to_soar_in_turbulent_environments}. There, the authors used DRL to train an algorithmic pilot
that should keep a glider in the air by taking advantage of thermic ascendant
currents. This is an application that may prove useful for increasing the
flight duration of light high-performance Unmanned Aerial Vehicles (UAVs). To
continue with UAVs, a recent work has highlighted the potential of DRL to
replace also their attitude control algorithms (usually implemented using a PID
controller or similar), which leads to increased flight performance and
maniability \citep{bohn2019deep}. Such enhanced performance using DRL algorithms has also been
observed when controlling quadcopter drones hovering in challenging
conditions, such as low-altitude translation where the drone interacts with its
own turbulent wake \citep{hwangbo2017control}. Finally, a recent work demonstrated the applicability of
DRL for high-level trajectory planning in complex environments where a
turbulent velocity map distorts the motion of an active drifter \citep{biferale2019zermelo}.

\begin{figure*}[ht]
\begin{center}
\includegraphics[width=.85\textwidth]{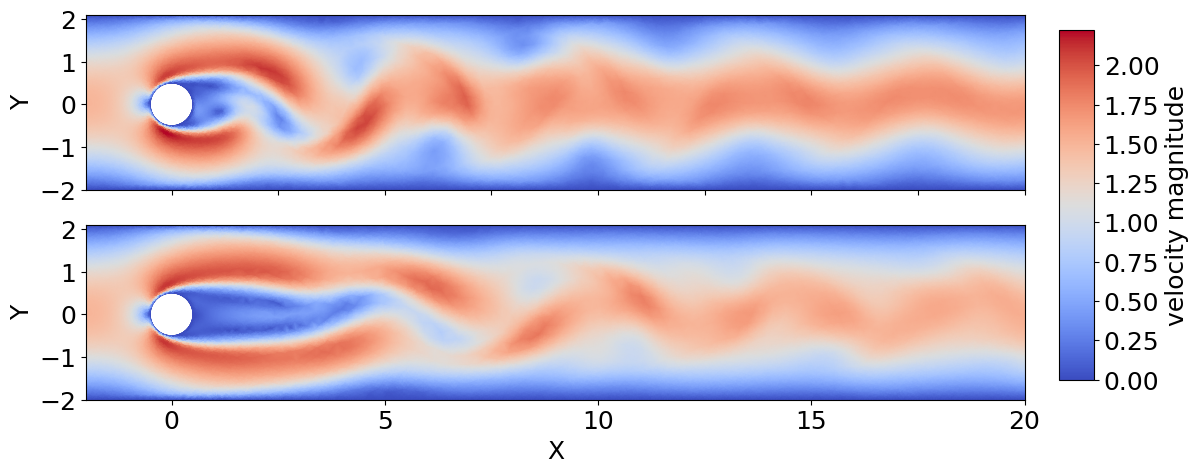}
    \caption{\label{fig:DRLControlJFM1} Illustration (velocity magnitude
    snapshot) of the effect of the
    active flow control strategy found by the PPO algorithm on the flow configuration
    behind a 2D cylinder at Reynolds number 100. The flow is altered in a
    way similar to what would be obtained with boat tailing, effectively reducing
    drag and lift fluctuations. Figure reproduced from \citet{rabault_kuchta_jensen_reglade_cerardi_2019}.}
\end{center}
\end{figure*}

\begin{figure*}[ht]
\begin{center}
\includegraphics[width=.45\textwidth]{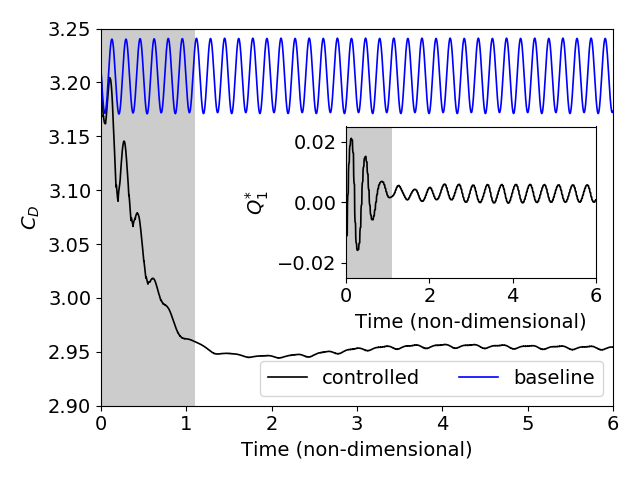}
    \caption{\label{fig:DRLControlJFM2} The control law applied by the PPO agent on
    the simulation presented in Fig. \ref{fig:DRLControlJFM1}. Two phases
    are clearly visible in the control law. The normalized mass flow rate $Q$ is only
    a few percents of the mass flow rate intersecting the cylinder. Figure reproduced from \citet{rabault_kuchta_jensen_reglade_cerardi_2019}.}
\end{center}
\end{figure*}

The second domain of application is interested in performing active control of
flow instabilities and separation. This, in turn, can lead to positive effects
such as reduced drag and reduced lift fluctuations in engineering designs.
Several recent works have, there also, highlighted the potential of DRL. The
first results served as a proof-of-concept of the methodology and illustrated
possible gains on a very simple flow configuration, namely, the 2D Karman alley
behind a cylinder at a moderate Reynolds number of Re=100 \citep{rabault_kuchta_jensen_reglade_cerardi_2019}. There, it was shown
that DRL could learn accurate control laws able to sensibly reduce drag by
using small jets on the top and bottom of the cylinder.
The effect of control on the flow configuration is illustrated
in Fig. \ref{fig:DRLControlJFM1}. Since the control can be applied at the location of the flow
separation, the mass flow rate necessary to change the flow configuration and
reduce drag can be as low as a fraction of a percent of the incoming mass flow
rate intersecting the cylinder. This example is illustrative of the potential
of DRL compared with other methods, such as the adjoint or parameter grid
search. Compared with the adjoint, only a limited insight into the system is
necessary to perform control, and as little as 5 probes are sufficient to
apply a sensible strategy. Moreover, once training has been performed,
obtaining the next control from a series of probes measurements is very
computationally inexpensive - as it is enough to just propagate the data of the
probes through the ANN. Compared with a grid search, which could for example
investigate the optimum jet forcing of the form $A cos(\omega t)$, where $A$ and
$\omega$ are the parameters to optimize, the DRL strategy is far more refined.
Indeed, the control law found by the DRL agent is composed of two phases:
first, a few cycles of relatively large actuations are performed to change the
configuration of the flow. By contrast, after the configuration of the flow has
been modified, the amplitude of the control is much reduced. This is illustrated
in Fig. \ref{fig:DRLControlJFM2}. The frequency of
the control also changes as time passes, and the final vortex shedding
frequency in the reduced drag configuration is modified compared with that of
the baseline flow.

Therefore, it appears that DRL can be used as an experimental tool to study the
properties of complex flows. Going to higher Reynolds numbers leads to more and
more complex dynamics, and underlines even better the adaptability of DRL
algorithms (ongoing work, see preliminary reports by \citet{FengAbsShang}). In particular,
DRL can effectively find control strategies that adapt to stochastic, noisy
systems where the closed-loop control allows to react to unforeseen dynamics. In order to control
more complex cases, it becomes important to speed up the training phase.
Luckily, this can be easily performed. Indeed, one can observe from Eqn. (5)
that the Monte Carlo sampling of trajectories in the phase space can be performed independently on the different
trajectories. This, in turn, means that several simulations can be run
completely independently of each other in a perfectly parallel manner for
collecting trajectories in the phase space. This results in large performance
gains in cases where the simulation of the environment is the limiting factor
in terms of computational resources, which is the case when CFD simulations are
used to train the DRL agent. Speedups of up to 60 times have been reported when
applying such a parallelization technique \citep{doi:10.1063/1.5116415}. In addition, this technique has the
potential to provide even larger performance gains for more challenging flows.
Indeed, the more complex the system to control, the more trajectories in the
phase space are needed to compute the policy gradient for the update of the
ANN.

Some additional important technical considerations are also presented in \citet{doi:10.1063/1.5116415}. In
particular, it is highlighted that the frequency of the action update by the
controller should be carefully adapted to the system to control. This is easily
understandable from simple physical considerations. Indeed, the control
frequency should be at least high enough to control the underlying phenomena,
which translates following the Nyquist criterion to the condition that $f_a > 2
f_s$, where $f_a$ is the frequency of the action update, and $f_s$ the
underlying typical frequency of the system to control. Similarly, the frequency
of the action update cannot be too high. Indeed, the exploration of the control
law by the DRL agent is based on trial-and-error. Therefore, if the frequency
of the action update is too high, the DRL algorithm will effectively force the
system to control with high frequency white noise at the beginning of training.
In this case, the probability that the white noise generated has some features
that perform some effective control on the system is very low, and the trial-and-error
attempts will likely fail to produce any progresses. By contrast, if the
frequency of the action update is slightly more moderate, the probability of a
`lucky attempt' that does affect the system in a positive way is much higher.
This leads to a sweet spot for the action update frequency, typically around
$f_a \approx 10 f_s$. If this frequency of update is different from the
numerical time step of the simulation, one needs to apply interpolation of the
control between action updates. In the case of a system presenting multiscale
properties, such a clear frequency $f_s$ cannot be easily defined. In this
case, it is possible to extend the output of the policy with one additional
value that lets the ANN choose also the duration until the next control update.

Another important point also highlighted in \citet{doi:10.1063/1.5116415} is that the DRL practitioner
should be careful in choosing a reward function that does not push the DRL
agent to perform degenerate, or `cheating', optimization. The ability of DRL
algorithms to take advantage of weaknesses of the environment to artificially
improve their reward in unexpected ways is well known from experimentations
with video games \citep{faultyboatreward, hideandseek}. Similar challenges are observed in the
case of flow control. For example, an effective reduction of the drag value can
be obtained by strong biased blowing upwards or downwards on the sides of a
cylinder, but in this case a large value of lift is produced at the same time
as drag is reduced due to the bending of the wake. However, this is not really
a strategy that performs flow control (or not only flow control). Therefore, 
adding a lift penalization to
the reward function is found to be necessary in order to observe effective,
unbiased flow separation control in the case of the 2D cylinder.

\begin{figure*}[ht]
\begin{center}
\includegraphics[width=.99\textwidth]{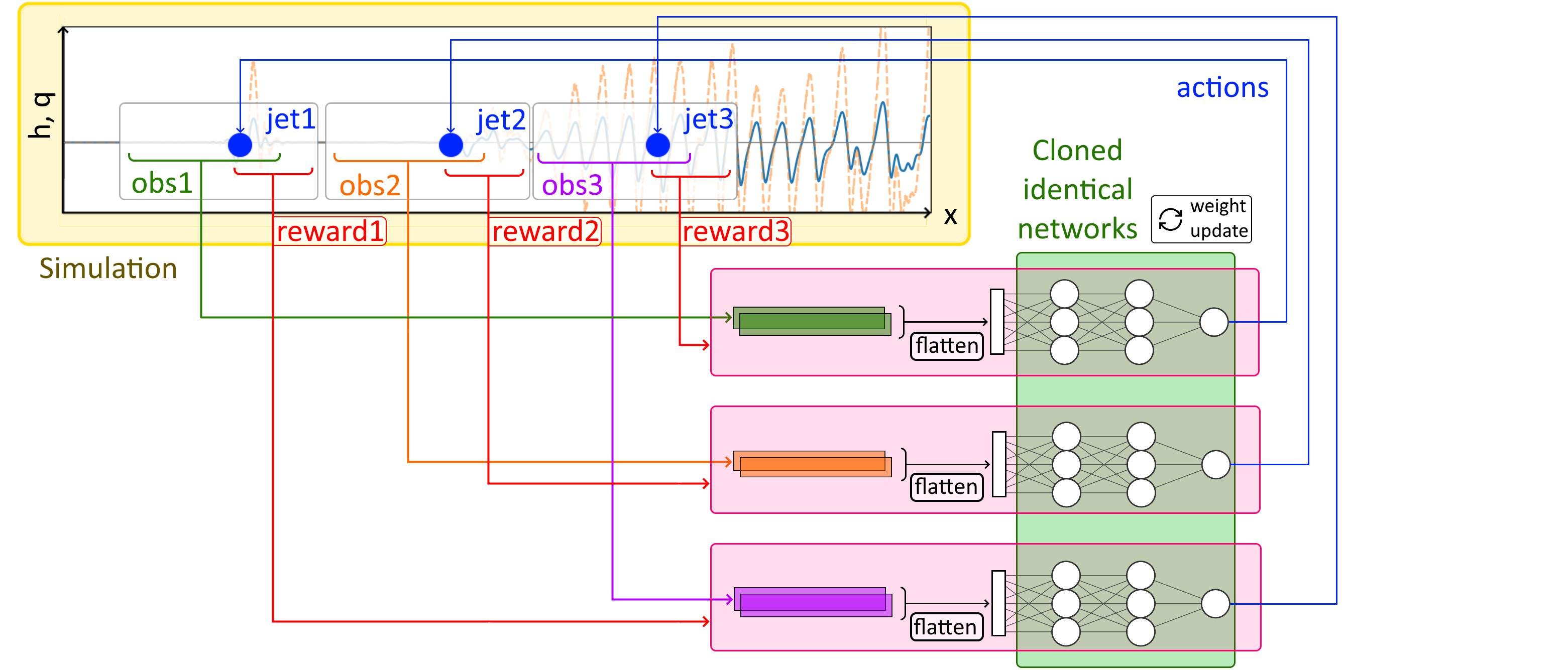}
    \caption{\label{fig:using_invariants} In the case when systems with large
    control space dimensionality are considered, exploration and training
    can reach prohibitive costs. However, such systems also usually feature some
    properties of locality and invariance. This can be exploited effectively, either
    by using convolutional networks, or by resorting to cloned ANNs as shown
    in this figure. By using cloned ANNs, it is possible to both apply the same strategy
    on all outputs, therefore reducing the cost of exploration, and to
    simultaneously increase the volume of reward generated, therefore,
    providing unambiguous signal during training and improving training
    quality. Figure reproduced from \citet{doi:10.1063/1.5132378}.}
\end{center}
\end{figure*}

In addition to these considerations about the application of DRL, the results
of \citet{doi:10.1063/1.5132378, nanjingconf2019}
illustrate the importance of using the locality and invariance properties of
the underlying environment in order to make learning more tractable. This is
especially true for systems that present a control domain of large
dimensionality. Indeed, as highlighted by \citet{doi:10.1063/1.5132378}, failing
to take these properties into account can lead to prohibitive exploration and
learning costs, effectively making DRL training impossible, even when quite
simple systems are considered. By contrast, satisfactorily reproducing locality
and invariance in the architecture of the ANN, for example through the use of
convolutional networks or by using cloned ANNs to control different parts of
the environment as shown in Fig. \ref{fig:using_invariants}, allows to
successfully discover effective control strategies
even when large systems are considered.

Following these promising results, several groups are now working towards
obtaining effective control laws in more challenging situations. These include,
pushing higher the Reynolds number in simulations similar to the 2D 
cylinder control of \citet{rabault_kuchta_jensen_reglade_cerardi_2019} (see the
preliminary presentations of \citet{FengAbsShang} for example on this aspect),
and performing control of more realistic 3D
flows. This approach should probably be
adopted also in future applications of DRL: first, one has to start with
a case of reduced complexity to apprehend some of the potential pitfalls of DRL
and how to mitigate them, before moving to the full complexity problem.

While the application of DRL to active flow control in simulations starts to be
well established and to diffuse in the fluid mechanics community (see the work
of other groups on different configurations and problems: control of a chaotic
system \citep{doi:10.1098/rspa.2019.0351}, control of Rayleigh-Benard convection \citep{corbetta2019reinforcement}), no physical experiments have been
presented yet, to the best of the knowledge of the authors,
in the context of Fluid Mechanics (of course, many applications have been presented
in the context of robotics and control of industrial systems). This may be related
to several challenges. First, building an experiment is nowadays more time
consuming and complex than setting up a simulation using an open source
simulation package. Second, applying DRL to experiments effectively sets
demanding requirements on the underlying electronics and computer
infrastructure. Indeed, the network prediction and training must be performed
within tight time constraints to follow up with the physical phenomena. While
this is in theory not a specially problematic difficulty, since the
availability of low-latency Real Time OSs, fast FPGAs, and effective neural
network accelerators such as GPUs is well established, in practice this demands
that experimental fluid mechanics teams extend their technical skillset which,
while not a fundamental challenge, may take some time. On this aspect, it would
be very useful if some groups with dedicated hardware competence
in the Fluid Mechanics or robotics community
could release designs of effective, GPU or FPGA-accelerated systems adapted
to the training of real-world controllers in a similar way as open source software
packages are released. Third, and maybe more
fundamental, it is usually more challenging to acquire large amounts of
real-time, high accuracy data within an experiment than in a simulation
environment. For example, the main training of \citet{rabault_kuchta_jensen_reglade_cerardi_2019} uses 151 velocity probes
located both around the cylinder and further in its wake. Performing such large
data acquisition in real time may be challenging even with high speed PIV in a
carefully crafted experiment, and it will definitely be very challenging, if
doable at all, in a realistic setup corresponding to an industrial application.
However, there also some solutions exist. In particular, \citet{rabault_kuchta_jensen_reglade_cerardi_2019} also shows that the
number of probes can be drastically reduced with only a limited decrease in the
control efficiency, at least in the situation investigated. Preliminary results
(ongoing work in our group) even indicate that, in a case similar to \citet{rabault_kuchta_jensen_reglade_cerardi_2019} aiming
at controlling the flow around a 2D cylinder at moderate Reynolds numbers, having as state
observation a few pressure probes located at the surface of the cylinder is
enough to derive a meaningful strategy. This, in turn, is very achievable in a
real-world setup. Finally, the last difficulty is that one needs to be able to implement physical actuators
that can follow up with the control strategy in order to apply the control to a
real world system. While there are many candidates for such actuators and the
literature usually considers that control algorithms, rather than physical
actuators, are the main limitation to performing active flow control \citep{collis2004issues, cattafesta2011actuators}, this
is yet another technical domain that needs to be mastered to perform a physical
experiment.

Therefore, ideally, the numerical simulations and the experiments should be
complementary of each other. The numerical simulations can be used as easily
set-up benchmark cases, to be tested and well understood before real-world
implementation. In addition, it is well established that ANNs can, given
suitable care is taken during training, be trained or at least pre-trained in
simulations before transferring the strategies learnt to the real world \citep{hwangbo2019learning} (see
\url{https://www.youtube.com/watch?v=aTDkYFZFWug&feature=youtu.be} for a nice illustration).
These abilities of transfer learning and mitigation of the reality gap between
simulations and the real world are one more strength for DRL compared with other
methodologies. The strength of experiments, by contrast with simulations, will be both to present
indisputable real-life validation of the strategies and methodologies revealed
by simulations, and to open the way to full scale Reynolds numbers applications.
Indeed, real-life experiments will be the only way to generate large amount of training
data in situations as costly to simulate as realistic full-scale flows.

\section{Application to shape optimization}

As a second possible domain of application for DRL within Fluid Mechanics
both shape optimization and, more generally, the workflow of a CFD engineer
designing a part can be considered. This is especially attractive
when
sophisticated geometries have to interact with complex nonlinear flow features,
which can be a challenge to adjoint-based methods due to the existence of many
local minima.

However, applications in this domain are still relatively less developed than
in the case of flow control. This may be explained in part due to the existence
of other well established non-gradient-based methods for these tasks, such as
evolution-based algorithms.  In addition, it is slightly less clear how DRL should be applied to such
problems. Indeed, one can either attempt to perform `one shot optimization'
where the ANN has to directly describe the shape it believes to be optimal, or
`iterative optimization' where the ANN is allowed to incrementally deform the
shape to make it evolve towards an optimal geometry. Since DRL is usually more
efficient in closed-loop setups, as this is how the algorithms were initially
thought of and designed, the second approach may be more promising on the long
term. However, implementing such iterative methods puts harder requirements on
for example the mesh deformation and topology checking, and as a consequence
the applications of DRL to shape optimization that the authors are aware of are
based on one shot optimization using parametric models.

\begin{figure*}[ht]
\begin{center}
\includegraphics[width=.95\textwidth]{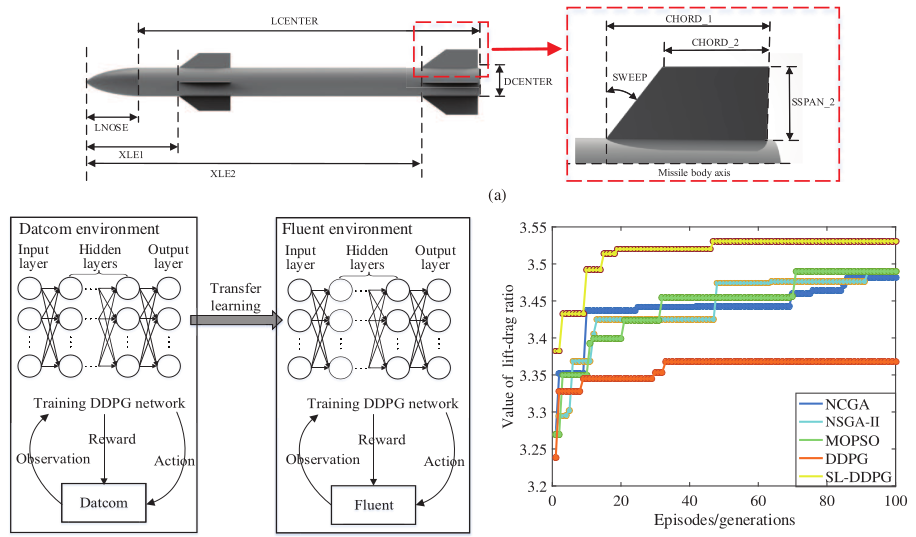}
    \caption{\label{fig:missile_optimization} Illustration of the optimization
    parameters used for maximizing the flight performance of a model missile.
    The results indicate that the tuned DDPG method
    is both competitive in terms of absolute performance as well as sample
    effective compared with traditional optimization methods.
    Figure reproduced from \citet{yan2019aerodynamic}.}
\end{center}
\end{figure*}

The first such application considers aerodynamic optimization of a missile-like
object flying within given specifications \citep{yan2019aerodynamic}. The DRL algorithm starts from a
reasonable missile shape, and is charged to perform limited changes to the
geometry of the missile in order to improve the flight performance. The
geometry changes are characterized through a series of design parameters such
as wing dimensions or missile length, as visible in Fig \ref{fig:missile_optimization}.
In this work, it is shown that while a
policy gradient DRL algorithm (in this case Deep Deterministic Policy Gradient, DDPG) is beaten by other
trial-and-error methods when applied as is out-of-the-box, a slightly tuned
version of this algorithm introduced by the authors (SL-DDPG) beats these other
techniques in terms of both speed of convergence and quality of the optimal
configuration found, as also visible in Fig. \ref{fig:missile_optimization}.

In the second application of DRL to shape optimization that the authors are
aware of \citep{viquerat2019direct}, the PPO algorithm is charged with creating a shape that maximizes
lift at moderate Reynolds numbers. By contrast to the first application, this
time the DRL algorithm has to perform one-shot optimization starting from a
geometry that does not produce any lift, in this case a cylinder discretized
with second-order splines. There also, it is found that the DRL algorithm
successfully performs optimization upon unambiguous parametrization of
the geometry to optimize by
splines, which is obtained by limiting the free domain of the defining points to separate
quadrants in order to avoid overlap and self-crossing of the shape boundary.
The parameters defining the splines are also re-normalized to a typical range
of $[-1;1]$, where the ANNs perform best. In addition, non-meshing shapes that
can be obtained for example by using too high local curvatures of the shape are
punished by applying a negative reward and immediately starting the next
optimization step. Following these additions to the methodology, the DRL
algorithm is there also able to produce shapes that generate large values of
lift. Unsurprisingly, the shapes obtained are wing-like, with an exact
morphology that depends on the number of spline points let free to be used in
the optimization.

While these works confirm that DRL also has some potential in shape
optimization, several additional improvements should be investigated in the
future. First, making the DRL algorithm aware of the details of the flow
configuration by giving it access to the flow fields should
help to build a high-level understanding of the underlying problem. Second, one
can expect that using a parametrization of the shape being optimized that is
not case-specific, for example by letting the ANN deform the mesh of the shape
rather than just change a few characteristic coefficients, would allow to offer
a more general optimization context and lead to some possibilities for
generalization and transfer learning, which are known strengths of ANN-based
algorithms. However, this may require additional complexity in the
implementation of the environments, since mesh morphing techniques are more
complex to implement than parametric shape descriptions.

\section{Future prospects and conclusions}

The interest in using DRL techniques for Fluid Mechanics applications is
rapidly growing, following the many successes of this technique in a wide range
of domains of science. This is for example visible through the number of
DRL-centered publications at the APS/DFD meeting. While in 2018 only 1 talk was
focused on DRL, in 2019 over 10 talks were specifically mentioning DRL in their
title (data collected by searching after Deep Reinforcement Learning keywords
in the titles of talks at the APS/DFD from the official websites: \url{http://meetings.aps.org/Meeting/DFD18/SearchAbstract} and
\url{http://meetings.aps.org/Meeting/DFD19/SearchAbstract}).

The reason for this interest lies in the potentialities of DRL to successfully
analyze situations that are challenging for more classical methods due to the
combination of non-linearity, non-convexity, and high dimensionality, that can be
found in many complex problems. The ability to perform well on
such challenging problems comes from the nature of DRL
algorithms. Indeed, since they are performing exploration based on a trial-and-error
approach, they are not easily trapped in suboptimal local extrema and are, therefore, 
well adapted to systems with nonlinear stochastic dynamics and partial observations. 

As a consequence, DRL is especially promising for at least two applications
within Fluid Mechanics that correspond well to its formal framework, namely
active flow control and shape optimization. So far, applications for both cases
have been performed on relatively simple benchmarks. However, several groups
are working towards the use of DRL in more challenging configurations, and the
good results obtained on the simplest cases are so far promising for further
application of this methodology. In addition, starting with simple
configurations allows to quickly test different ways to use DRL, and to gain
experience on the main points one should be careful about when trying to apply
DRL to a new problem. In particular, at least four important factors were
highlighted by the works performed so far. First, DRL can be relatively
data-hungry, but luckily the algorithm is naturally parallel which allows to
obtain large speedups given that enough computational resources are available.
Second, renormalization of data to a range close to $[-1; 1]$ where the
discontinuity of the neurons is active is key to the good functioning of these
methods. Third, the choice of the reward function is both of key importance for
the success of optimization, and sometimes challenging. Indeed, the DRL
algorithms are very efficient at finding flaws or shortcuts in the
environments, and therefore they easily optimize naive reward functions in
unexpected and unwanted ways. Fourth, the application of DRL to large systems
with many control values can become challenging due to the curse of dimensionality on
the control space dimension, which translates into prohibitive exploration
costs. Luckily, exploiting locality and invariants in the environment to
control allows for effective mitigation of this difficulty.

Following these progresses, it is now possible to envision the control of
higher Reynolds number, 3D flows that are closer to real-world systems. In
addition, the use of DRL in real-world experiments rather than simulations
appears more likely as the strengths of the method are revealed through
numerical benchmarks. Ultimately, one can hope that DRL will become both a tool
of practical importance for industrial applications, and a methodology that
allows to investigate general properties of flows through an empirical approach.

Finally, we note that the Fluid Mechanics community is by no ways alone in
investigating the use of DRL on complex physical or mechanical systems.
Arguably, many of the applications mentioned in this review are at the
interface between fluid mechanics, robotics, biomimetism, and computer science.
As a consequence, there is much knowledge and experience to gain from adopting
a multidisciplinary approach, at least when reading through the DRL literature.
This is particularly true when general questions around DRL are investigated,
such as the possibility for transfer learning or bridging the reality gap
between training in simulations and application of the policies found
thereafter to the real world. Similarly, it is probably important for the Fluid
Mechanicists to be careful to not ignore tools and software packages relased by
these other communities. For example, many excellent resources and
implementations are available open source around the PPO algorithm, and
it is probably a good idea to resort to these well-tested and benchmarked code
bases, and help provide feedback and improve them when necessary, rather than
re-developing from scratch new tools that largely re-implement the same
features. As a consequence, a key enabler to further DRL applications will
likely be the ability of the community to commit to further sharing code and
benchmarks as open source, so as to reduce the entry barrier for new groups who
want to participate in DRL research. On this aspects, the openness of the DRL
community, that has been widely sharing code and software packages, must be
acknowledged. We hope such openness can continue also thanks to Fluid
Mechanicists themselves, for example with the sharing of effective hardware
designs to make DRL easier to use in real-world experiments with hard time
constraints.

\bibliographystyle{jfm}
\bibliography{template}

\end{document}